\documentclass[pra,aps,twocolumn,superscriptaddress,showpacs,tightenlines,amsmath]{revtex4}
\usepackage{epsfig}
\usepackage{amsmath,amssymb}

\begin{document}

\title{Measurement in control and discrimination of entangled pairs under self-distortion}

\author{Francisco Delgado}
\email{fdelgado@itesm.mx}
\affiliation{Mathematics and Physics Department, Quantum Information Processing Group, Tecnologico de Monterrey, Campus Estado de Mexico, Atizapan, Estado de Mexico, CP. 52926, Mexico}

\date{\today}

\begin{abstract}
Quantum correlations and entanglement are fundamental resources for quantum information and quantum communication processes. Developments in these fields normally assume these resources stable and not susceptible of distortion. That is not always the case, Heisenberg interactions between qubits can produce distortion on entangled pairs generated for engineering purposes (e. g. for quantum computation or quantum cryptography). Experimental work shows how to produce entangled spin qubits in quantum dots and electron gases, so its identification and control are crucial for later applications. The presence of parasite magnetic fields modifies the expected properties and behavior for which the pair was intended. Quantum measurement and control help to discriminate the original state in order to correct it or, just to try of reconstruct it using some procedures which do not alter their quantum nature.

Two different kinds of quantum entangled pairs driven by a Heisenberg Hamiltonian with an additional inhomogeneous magnetic field which becoming self-distorted, can be reconstructed without previous discrimination by adding an external magnetic field, with fidelity close to 1 (with respect to the original state, but without discrimination). After, each state can be more efficiently discriminated. The aim of this work is to show how combining both processes, first reconstruction without discrimination and after discrimination with adequate non-local measurements, it's possible a) improve the discrimination, and b) reprepare faithfully the original states. The complete process gives fidelities better than 0.9. In the meanwhile, some results about a class of equivalence for the required measurements were found. This property lets us select the adequate measurement in order to ease the repreparation after of discrimination, without loss of entanglement. 
% Keywords: Quantum control; quantum information; Heisenberg model 


\end{abstract} 

\maketitle

\section{Introduction}

Quantum information processing has been a field of increasing activity in diverse areas as information theory, control, engineering and measuring \cite{nielsen1}. Physical elements of quantum information applications are bounded to physical imperfections as decoherence or even self distortion by interactions between their parts. Additional procedures have been introduced to correct these deviations from desired behavior. Quantum control deals with this kind of problems measuring, analyzing and providing feedback to the system in order to implement those procedures considering that its alteration upon measurement is not completely quantifiable. Quantum control was propelled by notable works of control in nanoscale systems \cite{milburn1}, in quantum feedback control \cite{doherty1, doherty2} and in quantum systems under continuous feedback \cite{wiseman1, clark1}. Some specific procedures of quantum control are based on exploiting the system properties in order to drive it without the use of projective or weak measurements \cite{mielnik1, fernandez1, delgado1}. Processes to introduce quantum control have been recently developed in order to discriminate states by taking measurements and using feedback  on single qubits \cite{wang1, yuang1, branczyk1, xi1} and on entangled qubits \cite{delgado2}.

Quantum correlations and entanglement are the basis for quantum computation and quantum communication. Normally, the stability of these items is assumed, but as almost any physical system that is not always true. Some recent experimental work shows how to produce entangled spin qubits in quantum dots and electron gases \cite{saraga1, saraga2}, in which, its identification and control are basic for later applications, because input variations in the initial configuration of the entangled pair are present. Related to this, \cite{delgado3} has presented some schemes of discrimination and repreparation to prevent the distortion which could emerge due to external non controllable interactions enabled by parasite fields which can distort the original state. In addition, some applications require precise knowledge of the produced state, so quantum measurement is necessary to characterize it without alter it.

In this paper, we show that both, discrimination and repreparation, are efficiently made by the previous reconstruction of the distorted state without discrimination, in agreement with the scenario and the set of prescriptions given in \cite{delgado3}. Discrimination is optimal for a class of generic equivalent measurements (ranging from local until non-local measurements) and its performing gives an efficient repreparation. Here, generic means independent from parameters of distortion. The paper is organized as follows: Section II explains reconstruction process related with results in \cite{delgado3}. Section III outlines the discrimination process and the kind of measurements involved under considerations in this paper. In section IV, equivalent measurements for the discrimination are discussed. In section V, we depict the repreparation of the original states after of their discrimination.

\section{Distortion control in Heisenberg interaction for bipartite qubits}

We are dealing with the problem depicted in Figure 1. A system generates known entangled states with the same probability, one state at a time out of two possible non-equivalent states. The user does not know which specific state is produced and in addition, some kind of distortion is introduced  because of the magnetic interaction between their parts before he has access to it. For quantum engineering purposes, this state is required to be identified and to be reconstructed into the original state in the best possible way. This is a simplification of the process presented in \cite{saraga1, saraga2}) and the control focus differs from that presented in \cite{delgado3} because here we are interested in both operations at time instead only one. Thus, we deal with an analogous system as showed in \cite{delgado3}, in which, an entangled pair in one of the two possible orthogonal states introduced there is generated:

\begin{eqnarray} \label{initialstates}
\left| \beta_1 \right> &=& \left| \beta_{00} \right> \nonumber \\ 
\left| \beta_2 \right> &=& \sin \theta \left| \beta_{01} \right> - \cos \theta \left| \beta_{10} \right> 
\end{eqnarray}

\noindent where $\left| \beta_{ij} \right>$ (for $i,j \in \{0,1 \}$) are the standard Bell states. As in \cite{delgado3}, $\theta$ is a parameter which provides a monotone distinction between the states and the trace distance \cite{fuchs1} for them is $\delta(\rho_1,\rho_2)=\frac{1}{2} \rm{Tr} |\rho_1-\rho_2|=\sin^2 \theta$, where $\rho_1=\left|\beta_1 \right> \left< \beta_1 \right|$ and $\rho_2=\left|\beta_2 \right> \left< \beta_2 \right|$. The last state, $\left| \beta_2 \right>$, goes from a similar state to $\left| \beta_1 \right>$ when $\theta=0$ until a very different one when $\theta=\pi/2$. Note that initially, both states are maximally entangled.

Immediately, a self-interaction ruled by the Heisenberg hamiltonian with a possible inhomogeneous magnetic field in some direction begins between the pair of particles:

\begin{equation} \label{hamiltonian}
H= -J \vec{\mathbf{\sigma}}_1 \cdot \vec{\mathbf{\sigma}}_2+B_1 {\sigma_1}_z +B_2 {\sigma_2}_z
\end{equation}

In agreement with \cite{delgado2, delgado3}, after some time $t$ those states become distorted in: 

\begin{eqnarray} \label{distortedstate}
\nonumber \left| \beta_1' \right> &=&e^{itj}(\cos (b_+ t')  \left| \beta_{00} \right> - i \sin (b_+ t')  \left| \beta_{10} \right>)  \\
\nonumber \left| \beta_2' \right> &=&(e^{-ijt'} (2ij \sin t' + \cos t') \sin \theta  \left| \beta_{01} \right> - \\ 
\nonumber & & e^{itj} \cos \theta \cos (b_+ t') \left| \beta_{10} \right>) + \\ 
\nonumber & & (i e^{ijt'} \cos \theta \sin (b_+ t') \left| \beta_{00} \right> - \\ 
\nonumber & & i b_-  e^{-ijt'} \sin \theta \sin t'  \left| \beta_{11} \right> ) \\
\end{eqnarray}

\noindent where: $b_+=(B_1+B_2)/R$, $b_-=(B_1-B_2)/R$, $j=J/R$, $R=\sqrt{(B_1-B_2)^2+4J^2}$ and $t'=Rt$. In the following we drop the prime in the time variable. This interaction is trace distance preserving: $\delta(\rho_1',\rho_2')=\sin^2 \theta$. The first state in (\ref{distortedstate}) remains maximally entangled always but not the second one; some entanglement is lost because the distortion. \cite{delgado3} shown two different control procedures for reprepare the original state, here we are selected the first on-site control procedure which is more accurate. In the following we refer to it as a reconstruction process instead of repreparation process as in that work, because we need differentiate it from the repreparation process after of a later discrimination. So, applying an extra homogeneous magnetic field during time $T$ after $t$, we obtain the complete evolution operator:

\begin{equation} \label{evolsit1}
U(t+T)=U_{b_+ + \delta b_+}(T)U_{b_+}(t)
\end{equation}

\noindent the first driving the distortion and the second one driving the reconstruction (in a greater average field $b_+ \rightarrow b_+ + \delta b_+$, with $b_-$ unchanged), in agreement with \cite{delgado2}. This last operation has the control parameters:

\begin{eqnarray} \label{solsit1}
T &=& n \pi-t \nonumber \\
\delta b_+ &=& \frac{\pi(2m-n(b_+ - 2j+1))}{T} \nonumber \\
\frac{s}{2n} &=& Q(j) \nonumber \\
{\rm with:} && n, m, s \in \mathbb Z
\end{eqnarray}

\noindent where $Q(j)$ is a rational approximation to $j$, with $2n$ as denominator. With a suitable selection of $n$ and $s$, we can have $Q(j)$ as close to $j$ as we want (in the case $j \in \mathbb Q$, $j=Q(j)$ is always possible). Thus, we obtain a quasi evolution loop (until unitary factors) $U(t+T)=I'$, with $I'$ the diagonal matrix: $I'={\rm diag}(1,1,1,e^{4 i n \pi \delta})$, where $\delta=j-Q(j)$. The reconstructed states become:

\begin{eqnarray} \label{reconstructedstate}
\nonumber \left| \beta_1'' \right> &=&(1+ie^{2in\pi \delta}\sin 2n\pi \delta)  \left| \beta_{00} \right> - i e^{2in\pi \delta}\sin 2n\pi \delta  \left| \beta_{10} \right>  \\
\nonumber \left| \beta_2'' \right> &=&\sin \theta  \left| \beta_{01} \right> - \\ 
\nonumber & & e^{2in\pi \delta} \cos \theta \cos 2n\pi \delta \left| \beta_{10} \right> + \\ 
\nonumber & & i e^{2in\pi \delta}\cos \theta \sin 2n\pi \delta  \left| \beta_{00} \right>  \\ 
\end{eqnarray}

Selection of $\cos 4\pi n \delta$ as close to $1$ as possible is desirable by means of a suitable $n$. In this sense, some restrictions related with the nature and the knowledge of $j$ are remarked in \cite{delgado3}. The reconstruction's average fidelity becomes:

\begin{eqnarray} \label{fidsit1}
F_{\rm N} &=& \frac{1}{2}\left(|\left<\beta_1|\beta_1'' \right>|^2 + |\left<\beta_2|\beta_2'' \right>|^2\right)= \nonumber \\
&=& 1-\frac{1}{2}\sin^2 2n\pi \delta (1+\cos^2 \theta)
\end{eqnarray} 

The reason of previous reconstruction is clear: take advantage of the same properties of magnetic field which generates the distortion in order to make a more feasible discrimination process than that obtained in \cite{delgado3}. Note that if no further discrimination and repreparation is made, then $F_{\rm N}$ is the Do-nothing process fidelity.

\section{Measurement problem for discrimination}

\subsection{General problem of measurement for discrimination}

The general problem of measurement for discrimination is given by a set of measurement operators $\{ M_i | i=1, ...,m \}$:

\begin{equation} \label{measureset}
\sum_{i=1}^m E_i =\sum_{i=1}^m M_i^\dagger M_i =1
\end{equation}

From this set of operators we identify two subsets which correspond to each qubit to be identified:

\begin{eqnarray} \label{eachq}
\mathcal{E}_1 + \mathcal{E}_2 &=&1\\ \nonumber
\mathcal{E}_k&=&\sum_{i \in \{i|f(i)=k \}}^m E_i  \\ \nonumber  {\rm with:} && k=1,2
\end{eqnarray}

\noindent where $f(i)=k$ assigns each measurement operator $E_i$ with some $\mathcal{E}_k$ in agreement with some predefined criteria. Then, the average fidelity is:

\begin{eqnarray} \label{fidav}
\Bar{F}&=& \frac{1}{2} \sum_{j=1}^2 \sum_{k=1}^2 P_{H_{k}}^{(j)}{\rm Tr}(\overline{\overline{\rho}}_j \rho_k)
\end{eqnarray}

\noindent where $\rho_k$, $\overline{\rho}_k$ and $\overline{\overline{\rho}}_k$ are the density matrices for desired, pre-measured and  reprepared states respectively. $P_{H_{k}}^{(j)}$ are the Helmstrom probabilities and their complements: 

\begin{eqnarray} \label{fidavb}
P_{H_{k}}&=&P_{H_{k}}^{(k)}={\rm Tr}(\overline{\rho}_k \mathcal{E}_k) \nonumber \\
1-P_{H_{k}}&=&P_{H_{k}}^{(j \ne k)}={\rm Tr}(\overline{\rho}_k \mathcal{E}_j)
\end{eqnarray}

\noindent for each qubit $k=1,2$. 

\subsection{Basis measurement transformations and fidelity invariance subgroup}

In order to discriminate the states presented in the last section, we need to make some measurement in order to decide if the premeasured state (here, distorted plus reconstructed state) came from $\left| \beta_1 \right>$ or $\left| \beta_2 \right>$. The objective of this subsection is to study some unitary transformation which can be made on the state previously to measurement, defining in this way an alternative measurement (Fig. 2):

\begin{eqnarray} \label{newmeasure}
M_i'=M_i U & \Rightarrow & E_i'={M_i'} ^\dagger M_i'=U ^\dagger E_i U \nonumber \\
& \Rightarrow & \mathcal{E}_k'=U ^\dagger \mathcal{E}_k U
\end{eqnarray}

We are especially interested on those measurements with the form:

\begin{equation}\label{es}
\mathcal{E}_k = \sum_{i,j \in s} \left| \phi_{ij} \right> \left< \phi_{ij} \right|
\end{equation}

\noindent where $s=\{(i,j)|2-\delta_{ij}=k; i,j=0,1 \}$ and $\mathcal{B}=\{ \left| \phi_{ij} \right> \ | i,j=0,1 \}$ is a complete orthogonal bipartite basis, not necessarily separable in general. The interest in them is because they are the experimentally simplest measurements. In particular, trough this work we will use the three measurement basis (with element definitions as in (\ref{es}) in that order):

\begin{eqnarray} \label{mbasis}
\mathcal{B}_{\rm C}&=&\{ \left| 00 \right > , \left| 01 \right > , \left| 11 \right >, \left| 10 \right >  \} \\
\nonumber \\
\mathcal{B}_{\rm B}&=&\{ \left| \beta_{00} \right >, \left| \beta_{01} \right > , \left| \beta_{10} \right > , \left| \beta_{11} \right > \} \\
\nonumber \\
\mathcal{B}_{\rm R}&=&\{ \left| \rho_{00} \right >=\left| \beta_{00} \right > , \nonumber \\
& & \quad \left| \rho_{01} \right >=\sin \theta \left| \beta_{01} \right > -\cos \theta \left| \beta_{10} \right >, \nonumber \\
& & \quad \left| \rho_{10} \right >=-\cos \theta \left| \beta_{01} \right > -\sin \theta \left| \beta_{10} \right >, \nonumber \\
& & \quad \left| \rho_{11} \right >=\left| \beta_{11} \right > \} \\ \nonumber
\end{eqnarray}

Note that there is an implicit correspondence between the elements trough of the three basis, which will be important below. We will interested in the unitary transformations $U$ which leave the fidelity invariant. In spite of (\ref{fidav}) and (\ref{fidavb}):

\begin{eqnarray} \label{invcond}
&  P^{(j)}_{H_k} & = \left < \beta_k' \right | \mathcal{E}'_j \left | \beta_k' \right > \\ \nonumber
& & = \left < \beta_k' \right | U \mathcal{E}_j U^\dagger \left | \beta_k' \right > \\ \nonumber
& & = \left < \beta_k' \right | \mathcal{E}_j \left | \beta_k' \right > \\ \nonumber
& & \Leftrightarrow U \mathcal{E}_j=\mathcal{E}_j U \\ \nonumber
\end{eqnarray}

\noindent so the strong condition is that $[\mathcal{E}_j,U]=0$, it means that $\{ \mathcal{E}_j \}$ is invariant. Imposing the last commuting condition on a general $U$ expressed in an arbitrary basis $\mathcal{B}$ like before, we obtain by direct calculation (note that $[\mathcal{E}_1,U]=0$ implies $[\mathcal{E}_2,U]=0$):

\begin{eqnarray} \label{ucomute}
U&=& \left(
\begin{array}{cccc}
a_{1,1}    & 0        & 0        & b_{1,1}  \\
0          & a_{2,1}  & b_{2,1}  & 0       \\
0          & b_{2,2}  & a_{2,2}  & 0       \\
b_{1,2}    & 0        & 0        & a_{1,2} 
\end{array}
\right) \\ \nonumber \\
{\rm with:} && \nonumber \\
&& a_{i,j}^2+b_{i,j}^2=1 ; \quad i,j=1,2 \nonumber \\
&& \sum_{k,l=1 \atop k \ne l}^2 a_{i,k}b_{i,l}=0; \quad i=1,2 \nonumber
\end{eqnarray}

By direct calculation we can show that these special transformations are a subgroup of the unitary transformations. Two special cases are the transformations (which will be useful for our discussion), where $\mathcal{H}$ is the Hadamard gate operator and ${\mathcal C}^1_2$ is the controlled-not gate:

\begin{eqnarray} \label{ubell}
U_{\rm CB}&=& {\mathcal C}^1_2 \mathcal{H}_1 {\mathcal C}^1_2 \nonumber \\
&=& \frac{1}{\sqrt{2}} \left(
\begin{array}{cccc}
1    & 0        & 0        &  1  \\
0          & 1  &  1  & 0       \\
0          & 1  & -1  & 0       \\
1    & 0        & 0        & -1 
\end{array}\right) 
\end{eqnarray}

\noindent (obviously expressed in the computational basis) which transforms the set:

\begin{eqnarray} \label{sstd}
\mathcal{M}_{\rm C}&=&\{ \mathcal{E}_{C_1}=\left| 00 \right > \left< 00 \right|+\left| 11 \right > \left< 11 \right|, \nonumber \\
&& \quad \mathcal{E}_{C_2}=\left| 01 \right > \left< 01 \right|+\left| 10 \right > \left< 10 \right| \} 
\end{eqnarray}

\noindent into the equivalent set:

\begin{eqnarray} \label{sbell}
\mathcal{M}_{\rm B}&=&\{ \mathcal{E}_{B_1}=\left| \beta_{00} \right > \left< \beta_{00} \right|+\left| \beta_{10} \right > \left< \beta_{10} \right|, \nonumber \\
&& \quad \mathcal{E}_{B_2}=\left| \beta_{01} \right > \left< \beta_{01} \right|+\left| \beta_{11} \right > \left< \beta_{11} \right| \} 
\end{eqnarray}

And:

\begin{eqnarray} \label{urot}
U_{\rm BR}&=&  \left(
\begin{array}{cccc}
1    & 0        & 0        & 0  \\
0          &  \sin \theta  & -\cos \theta  & 0       \\
0          & -\cos \theta  & -\sin \theta  & 0       \\
0    & 0        & 0        & 1 
\end{array}\right) 
\end{eqnarray}

\noindent expressed in the $\mathcal{B}_{\rm B}$ basis, which transforms:

\begin{eqnarray} \label{sbellp}
\mathcal{M}_{\rm B'}&=&\{ \mathcal{E}_{B'_1}=\left| \beta_{00} \right > \left< \beta_{00} \right|+\left| \beta_{11} \right > \left< \beta_{11} \right|, \nonumber \\
&& \quad \mathcal{E}_{B'_2}=\left| \beta_{01} \right > \left< \beta_{01} \right|+\left| \beta_{10} \right > \left< \beta_{10} \right| \} 
\end{eqnarray}
 
\noindent in:

\begin{eqnarray} \label{srot}
\mathcal{M}_{\rm R}&=&\{ \mathcal{E}_{R_1}=\left| \rho_{00} \right > \left< \rho_{00} \right|+\left| \rho_{11} \right > \left< \rho_{11} \right|, \nonumber \\
&& \quad \mathcal{E}_{R_2}=\left| \rho_{01} \right > \left< \rho_{01} \right|+\left| \rho_{10} \right > \left< \rho_{10} \right| \} 
\end{eqnarray}

\noindent note that $U^\dagger_{\rm CB}=U_{\rm CB}$ and $U^\dagger_{\rm BR}=U_{\rm BR}$.

\subsection{Measurement implementation}

Now, we will describe how to implement the last set of measurements. For the measurements in the computational basis $\mathcal{B}_{\rm C}$, Fig. 3a shows the single measuring of spin in $z$ direction on each particle. Fig. 3b shows the previous application of transformation $U_{\rm CB}$ followed by the measurement in the computational basis $\mathcal{B}_{\rm C}$ in the same order of identification which was mentioned after of formula (\ref{mbasis}). In some sense, this transformation is equivalent to the measurement in the nonlocal $\mathcal{B}_{\rm B}$ basis. Note that an additional application of the inverse transformation $U_{\rm CB}$ on the measured state (not included in the figure) gives the correct measured state in the Bell basis, nevertheless as a further process of reconstruction follows, this inverse transformation is skipped now. Finally, Fig. 3c shows the same procedure as in Fig. 3b but using ancilla qubits to make the measurements by entangling them with the qubits actually being discriminated (in the sense of the Neumark's theorem \cite{watrous1}). This last measurement process is included here only as reference of a cleaner process which does not touch the qubits of interest.

In agreement with the last section, the use of the measuring set $\mathcal{M}_{\rm C}$ or $\mathcal{M}_{\rm B}$ is completely equivalent for the fidelity of the process (assuming that an additional reconstruction on the measured state into another same state could be made for both cases). This sets are included in this work to discuss that despite in \cite{delgado3} an intuitive basis for discrimination was $\mathcal{M}_{\rm C}$, an additional attempt to use nonlocal measurements based on Bell basis and in $\mathcal{M}_{\rm B}$ does not give improving results for the fidelity.

Similarly, if the measuring set $\mathcal{M}_{\rm B'}$ is selected for the discrimination by using the last equivalent procedure with nonlocal measurement on Bell basis, it will be equivalent in the sense of fidelity invariance to the process using $\mathcal{M}_{\rm R}$ (but not with $\mathcal{M}_{\rm C}$ or $\mathcal{M}_{\rm B}$), in the order of identification given in (\ref{mbasis}), so an additional procedure based on the $U_{\rm BR}$ transformation, to measure in that basis with this set is not necessary. The importance on the implementation of these measurements will be explained in the following sections. The Figure 4 depicts the relations between the basis of measurement, the measurement operators and the fidelity invariance transformations discussed before.

\section{Optimum practical measurements and fidelities}
 
In the present section, we will to discuss some useful measurement for our control problem. At this point, we have the reconstructed states (\ref{reconstructedstate}) and then qubits are set far away so Heisenberg interaction stops, but other local or non-local operations can be still applied on them. The actual states form suggests the use of non-local measurements in the Bell basis. Nevertheless, it's well know \cite{presk1} that the optimal POVM to discriminate both states is:

\begin{eqnarray} \label{optpovm}
\mathcal{E}_k &=& M_k^\dagger M_k \nonumber \\
M_{k \ne k'}&=&\frac{1-\left| \beta_{k'} '' \right> \left< \beta_{k'} '' \right|}{1+| \left< \beta_k '' | \beta_{k'} '' \right>|}, \quad k,k'=1,2 \nonumber \\
\mathcal{E}_{\rm inc} &=& 1-\mathcal{E}_1-\mathcal{E}_2 
\end{eqnarray}

\noindent where $\mathcal{E}_{\rm inc}$ is stated for measurements which are inconclusive for discriminate (note that $\left< \beta_{k'}'' | \mathcal{E}_{\rm inc} | \beta_{k'}'' \right>=|\left< \beta_1'' | \beta_2'' \right>|$ are equal for $k'=1,2$). Nevertheless, as in this situation both states remain orthogonal, this is equivalent to take for $M_k$: 

\begin{eqnarray} \label{optimalbasis}
M_{{\rm op}_{k}}&=&\left| \beta_k '' \right> \left< \beta_k '' \right|, \quad k=1,2 \nonumber \\
\end{eqnarray}

Unfortunately this measurement basis is not experimentally practical always because it depends on the parameters of reconstruction $n$ and $\delta$.

\subsection{Bell basis measurements and optimum measurement}

As first approximation to the problem of find a practical basis of measurement, we propose a kind of measurements of the form:

\begin{equation} \label{nonlocal}
M=\sum_{i,j=0}^1 \alpha_{{ij}} \left | \beta_{ij} \right > \left < \beta_{ij} \right |
\end{equation}

Solving the optimization problem for $\alpha_{{ij}}$ parameters in order to maximize $P=\left< \beta_{k'} '' | M^\dagger M | \beta_{k'} '' \right>$ for each $k'=1,2$, a direct calculation gives the critical $P$'s for $k'=1$:

\begin{eqnarray} \label{m1}
M_{1_a}&=&\left | \beta_{00} \right > \left < \beta_{00} \right | \Rightarrow P_{1_a} = \cos^2 2n \pi \delta \nonumber \\
M_{1_b}&=&\left | \beta_{10} \right > \left < \beta_{10} \right | \Rightarrow P_{1_b} = \sin^2 2n \pi \delta 
\end{eqnarray}

\noindent and for $k'=2$:

\begin{eqnarray} \label{m2}
M_{2_a}&=&\left | \beta_{00} \right > \left < \beta_{00} \right | \Rightarrow P_{2_a} = \sin^2 2n \pi \delta \cos^2 \theta \nonumber \\
M_{2_b}&=&\left | \beta_{10} \right > \left < \beta_{10} \right | \Rightarrow P_{2_b} = \cos^2 2n \pi \delta \cos^2 \theta \nonumber \\
M_{2_c}&=&\left | \beta_{01} \right > \left < \beta_{01} \right | \Rightarrow P_{2_c} = \sin^2 \theta
\end{eqnarray}

As is expected, they depend on $n$ and $\delta$ and there are a conflict with both sets of measurements because depending on that values, the assignation could be made to $\left | \beta_1'' \right >$ or $\left | \beta_2'' \right >$ discrimination. Here, we will assume that we are in the $\cos 4\pi n \delta=1$ regime, so is more convenient to assign $\left | \beta_{00} \right > \left < \beta_{00} \right |$ for the discrimination of $\left | \beta_1'' \right >$ and $\left | \beta_{10} \right > \left < \beta_{10} \right |$ for the discrimination of $\left | \beta_2'' \right >$. Note that $\left | \beta_{11} \right > \left < \beta_{11} \right |$ is not only inconclusive for both cases but $\left < \beta_{11} | \beta_{k'}'' \right >=0$, so its measurement operator can be included in $\mathcal{E}_{\rm inc}$ or instead in $\mathcal{E}_1$ or $\mathcal{E}_2$ without effect. For simplicity in the notation used in section II, we include it in $\mathcal{E}_1$. It means that this is equivalent to use $\mathcal{M}_{\rm B'}$ as measurement operators for discrimination.

Nevertheless that the use of $\mathcal{B}_B$ in (\ref{nonlocal}) is not general, there are an argument about the practicity to use the last measurement operators. Extending the set (\ref{optimalbasis}) onto two other orthogonal operators $\left| \beta_k '' \right> \left< \beta_k '' \right|$ for $k=3,4$. This problem can be solved by obtain $1-M_{{\rm op}_1}^\dagger M_{{\rm op}_1}-M_{{\rm op}_2}^\dagger M_{{\rm op}_2}$, then diagonalizing and calculating its eigenvectors. This subspace is degenerated so the selection is not unique. A convenient selection is constructed with:

\begin{eqnarray} \label{b34}
\left | \beta_3'' \right > &=&  -e^{-4 i n \pi \delta} \cos \theta \left | \beta_{01} \right > \nonumber \\
& & + i  e^{-2 i n \pi \delta} \sin 2 n \pi \delta \sin \theta \left | \beta_{00} \right > \nonumber \\
& & - e^{-2 i n \pi \delta} \cos 2 n \pi \delta \sin \theta \left | \beta_{10} \right > \nonumber \\
\left | \beta_4'' \right > &=& \left | \beta_{11} \right >
\end{eqnarray}

The interest aspect is that in the $\cos 4\pi n \delta=1$ regime, this set of measurements reduces exactly to $\mathcal{M}_{\rm R}$ which is equivalent to $\mathcal{M}_{\rm B'}$. 

\subsection{Fidelities}

\begin{table} \label{tabla1}
\centering
\begin{tabular}[b]{|c|c|c|}
\hline
{\bf Basis} & $\left| \beta_{1}'' \right>$ & $\left| \beta_{2}'' \right>$ \\
\hline
$\left| 00 \right>$ & $\frac{1}{2}$ & $\frac{1}{2} \cos^2 \theta$  \\
$\left| 01 \right>$ & $0$ & $\frac{1}{2} \sin^2 \theta$  \\
$\left| 10 \right>$ & $0$ & $\frac{1}{2} \sin^2 \theta$  \\
$\left| 11 \right>$ & $\frac{1}{2}$ & $\frac{1}{2} \cos^2 \theta$  \\
\hline
$\left| \beta_{00} \right>$ & $\cos^2 2n \pi \delta$ & $\cos^2 \theta \sin^2 2n \pi \delta$  \\
$\left| \beta_{01} \right>$ & $0$ & $\sin^2 \theta$  \\
$\left| \beta_{10} \right>$ & $\sin^2 2n \pi \delta$ & $\cos^2 \theta \cos^2 2n \pi \delta$  \\
$\left| \beta_{11} \right>$ & $0$ & $0$  \\
\hline
$\left| \rho_{00} \right>$ & $\cos^2 2n \pi \delta$ & $\cos^2 \theta \sin^2 2n \pi \delta$  \\
$\left| \rho_{01} \right>$ & $\cos^2 \theta \sin^2 2n \pi \delta$ & $\cos^2 \theta \cos^2 2n \pi \delta (1+\sin^2 \theta)+\sin^4 \theta$  \\
$\left| \rho_{10} \right>$ & $\sin^2 \theta \sin^2 2n \pi \delta$ & $\cos^2 \theta \sin^2 \theta \sin^2 2n \pi \delta$  \\
$\left| \rho_{11} \right>$ & $0$ & $0$  \\
\hline
\end{tabular}
\caption{Measuring probabilities for each element of basis $\left| \phi_{ij} \right>$ discussed and for each reconstructed state $\left| \beta_k'' \right>$.}
\end{table}

We can analyze the Helmstrom probabilities calculating $|\left< \phi_{ij} | \beta_k'' \right>|^2$ for each basis element of $\left| \phi_{ij} \right>$ and for each reconstructed state $\left| \beta_k'' \right>$ (Table I). With them is easy calculate the Helmstron probabilities for each measurement operator set $\mathcal{M}$ and finally the final fidelity. For $\mathcal{\rm C}$ and $\mathcal{\rm B}$:

\begin{eqnarray} \label{fidCB}
F_{\rm C}=F_{\rm B}=1-\frac{1}{2}\cos^2 \theta
\end{eqnarray}

And for $\mathcal{\rm B'}$ and $\mathcal{\rm R}$:

\begin{eqnarray} \label{fidBR}
F_{\rm B'}=F_{\rm R}=1-\frac{1}{2}\sin^2 2n\pi \delta (1+\cos^2 \theta)=F_{\rm N}
\end{eqnarray}

It was expected that no best fidelity of $F_{\rm N}$ will be obtained. The only advantage is that in this process we has discriminated the states without additional cost (of course if we are able to reprepare the states faithfully  after of discrimination). The last fidelities are compared in the Figure 5. It is noticeable that near from the $\cos 4\pi n \delta=1$ regime and independently of $\theta$, fidelity is practically $1$. Far away from this regime the effects of the specific value of magnetic field and $\theta$  mentioned in \cite{delgado3} are present trough the imperfect reconstruction after of the distortion. Note specially that near from the case $\theta \approx 0$, $\mathcal{M}_{\rm C}$ or $\mathcal{M}_{\rm B}$ measuring operators become a better option instead $\mathcal{M}_{\rm B'}$ or $\mathcal{M}_{\rm R}$.

\subsection{Experimental limitations}

In \cite{delgado3} was mentioned that the unknowledge of $j$ can restrict the control effectiveness. The problem is that the term $n \delta$ in our expressions is not limited to go onto zero in a controllable way in order to make that $F \rightarrow 1$. While better $Q(j)$ will be to go $j$, normally $n$ is greater. By writing $j$ (assuming $N$ integer digits) as a succession of tenth powers: $j=\sum_{i=-N}^\infty a_i 10^{-i}$, then we obtain by taking as rational approximation of order $k$ only the first $k$ decimal digits in $j$: $n \delta=n (j-\sum_{i=-N}^k a_i 10^{-i})=10^k \sum_{i=k+1}^\infty a_i 10^{-i} < (a_{k+1}+1)/10 <1$. But, actually this is a shy limit because we assume that $n=10^k$, nevertheless if the $Q(j)=s/2n$ as in (\ref{solsit1}) is reducible, then the last limit is lower. If $j$ is a number with an arbitrary number of digits, but our actual knowledge of it is just to the $k$-digit, we can still use a worse rational approximation $Q(j)$ in order to $n \delta$ becomes lower by seeking the adequate $s$ and $n$ which make $Q(j)$ reducible. To illustrate this, we seeking numerically the best values of $Q(j)$ until $k=5$ digits for each $j \in (0,1]$ (note that $j>1$ gives the same results for $\sin^2 2 n \pi \delta$ in (\ref{fidBR})). Results are shown in the Figure 6. The graphic shows the best value $1-F_{\rm max}$ (when $\theta=0$) versus $j$ that is possible to find until $k=5$ digits of approximation for $Q(j)$. Each point depicts that solution and the whole graphic suggest the density of cases for each $1-F_{\rm max}$ value. This density is clearly lower for greater values of $1-F_{\rm max}$. Below, the cumulative distribution function, $\Omega$, corresponding to each $1-F_{\rm max}$ value. Note that around of 90\% of cases have up of 0.8 in fidelity.

\section{Final repreparation process}
 
In this section we will analyze the final process of repreparation after of the measurement and discrimination procedures studied before. In all cases we will assume that the measurement was make in the $\left| ij \right>$ basis, after of the appropriate $U$ transformation to induce an equivalent non-local measurement as was studied in the before sections.

\subsection{Basic gates for repreparation}
In the present section we will present different quantum computational gates which serve to reprepare the state emerging from measurement (which is always possible, see by example \cite{leung1}). In addition to ${\mathcal C}^1_2$ gate (which we do not discuss here), the repreparation process from the measured states induced by $\mathcal{M}_{B'}$ depicted in IIIC will require the local gate $U_{\theta}=I \cos \theta - i Z \sin \theta$ which can be physically generated by the qubit interaction with a $z$-magnetic field $B_z$ during $\tau$ time:

\begin{eqnarray} \label{ubz}
U_{B_z \tau}&=&  \left(
\begin{array}{cccc}
e^{-i B_z \tau}  & 0  \\
0  & e^{i B_z \tau}  
\end{array}\right) 
\end{eqnarray}

\noindent so $U_{\theta}=U_{B_z \tau}$ with $B_z \tau=\theta$. An additional necessary gate is the phase gate $S$, which can be obtained from this last until a unitary factor: 

\begin{eqnarray} \label{ese}
S &=&  \left(
\begin{array}{cccc}
1  & 0  \\
0  & i  
\end{array}\right) 
\end{eqnarray}

\noindent by selecting $B_z \tau=\frac{4p+1}{4}\pi, p \in \mathbb Z$. Hadamard gate can be generated until a unitary factor trough the similar gates to $U_{B_z \tau}$, but using magnetic fields in other directions:

\begin{eqnarray} \label{ubxby}
U_{B_\sigma \tau}&=& I \cos B_\sigma \tau - i \Sigma \sin B_\sigma \tau
\end{eqnarray}

\noindent with $\sigma=x, y, z; \Sigma=X, Y, Z$ respectively. With this, the Hadamard gate becomes (until a unitary factor):

\begin{eqnarray} \label{hadamard}
{\mathcal H}=U_{B_x \tau=\frac{2p+1}{2}\pi}U_{B_y \tau'=\frac{4q+1}{4}\pi}
\end{eqnarray}

\noindent with $p, q \in \mathbb Z$. Clearly, the operators $X, Y$ and $Z$ can be obtained (until unitary factors) respectively from (\ref{ubz}, \ref{ubxby}) by an adequate selection of $B_i \tau=\frac{2p+1}{2} \pi$ with $i=x,y,z$ and $p \in \mathbb Z$.

\subsection{Final repreparation}
After of discrimination, based on the knowledge of the original state (with sufficient certainty) we can apply a set of transformations in order to recover the original state, in agreement with the fidelity (\ref{fidCB}). The central aspects here is to note that ${\mathcal H}U_\theta {\mathcal H} \left| j \right>=\cos \theta \left| j \right>-i \sin \theta \left| j \oplus 1 \right>$, where $j=0,1$ and $\oplus$ is the sum module two; and ${\mathcal C}^1_2 {\mathcal H} {\mathcal C}^1_2$ is the transformation between computational basis and Bell basis depicted in section III. With that, one can easily reprepare the original state into the discriminated state by the set of measurements $\mathcal{M}_{\rm B'}$ in (\ref{sbellp}), beginning from the each one of the measured states, $\left| ij \right>$ obtained trough scheme depicted in the Figure 3b (or Figure 3c):

\begin{equation} \label{repreparation}
\left| \beta_{j + 1} \right>={\mathcal U}_{j + 1,ij} \left| ij \right>
\end{equation}

Table II shows until unitary factors, the precise transformations obtained from basic local and non-local transformations in the last subsection. \\

\begin{table} \label{tabla2}
\centering
\begin{tabular}[b]{|c|c|c|}
\hline
$\left| ij \right>$ & ${\mathcal U}_{j + 1,ij}$ & $\left| \beta_{j + 1} \right>$ \\
\hline
$\left| 00 \right>$ & ${(\mathcal C}^1_2 {\mathcal H}_1 {\mathcal C}^1_2)$ & $\left| \beta_{1} \right>$  \\
$\left| 01 \right>$ & ${(\mathcal C}^1_2 {\mathcal H}_1 {\mathcal C}^1_2) ( Y_1 S_1 ) ({\mathcal H}_1 U_ {\theta_1} {\mathcal H}_1)$ & $\left| \beta_{2} \right>$  \\
$\left| 11 \right>$ & ${(\mathcal C}^1_2 {\mathcal H}_1 {\mathcal C}^1_2) ( Y_1 S_1 X_1 ) ({\mathcal H}_1 U_ {\theta_1} {\mathcal H}_1)$ & $\left| \beta_{2} \right>$  \\
\hline
\end{tabular}
\caption{Transformations for the repreparation process for each one measured state identified with its discriminated state.}
\end{table}

\section{Conclusions}

The results of this work improving the control introduced in \cite{delgado2} by apply a general reconstruction before discrimination. After of this last, an additional process of repreparation give an almost perfect recovering of the original state depending upon the precise knowledge of $j$ coupling constant, which appears as the most delicate element in the whole process. The use of equivalent measurement basis make experimentally easier the use of appropriate non-local measurement to discriminate adequately the state after of reconstruction. Obviously, some details about the spatial management of pair of qubits are disregarded. Future work should be directed towards improving the control dependence on $j$ and their control of spatial position, by example using ion traps as in \cite{fernandez1}, because this is an central aspect in the whole control of the pair.

\section*{Acknowledgements}
I gratefully acknowledge Dr. Sergio Martinez-Casas for reviewing of this manuscript and Prof. Carlos Prado for fruitful discussions about density of rational numbers. 

\section*{Figure captions}

{\bf Figure 1}
Graphic description of the complete process of discrimination and repreparation including the ''Heisenberg distortion''. The system generates one out of two possible non equivalent bipartite entangled states. Then, during time $t$, the magnetic interaction introduces an internal distortion. The user apply for some time $T$ a reconstruction procedure still under Heisenberg interaction in order to reduce the possibility of error in the discrimination. After, some kind of convenient measurement is applied to discriminate the original state and finally some adequate repreparation process to recover the original state.

{\bf Figure 2}
Quantum circuit containing a unitary transformation $U$ preceding the measurement and additional transformation $U'$ following (by example, used for repreparation). The whole effect is to define a new measurement $M'_i=M_i U$. Note that the measurement device includes both qubits indicating a possible non-local measurement.

{\bf Figure 3}
Quantum circuits containing different measurement implementations. a) Basic measurement in the computational basis, b) transformation for make non local measurement in the Bell basis, and c) implementation of Bell non local measurements with ancilla qubits.

{\bf Figure 4}
Relations between the fidelity invariance transformations relating different measurement basis, which leave invariant and related some measurement operators.

{\bf Figure 5}
Comparison between $F_{\rm C}=F_{\rm B}$ (black) and $F_{\rm B'}=F_{\rm R}$ (chessboard layer) fidelities.

{\bf Figure 6}
Best values for $1-F_{\rm max}$ for each $j \in (0,1]$ by selecting the best $Q(j)$ until $k=5$ digits which minimizes $n \delta$. Below, the corresponding cumulative distribution function, $\Omega$.

\small  % Use 9 point text.

\end{document}